\documentclass[journal]{IEEEtran}

\pdfoutput=1

\usepackage{lineno}
\usepackage{graphicx}
\usepackage[noblocks]{authblk}
\usepackage{multirow}
\usepackage{url}
\usepackage{subfigure}

\begin{document}



\title{Prototype performance studies of a Full Mesh ATCA-based General Purpose Data Processing Board}


\author{Yasuyuki~Okumura, 
  Jamieson~Olsen, 
  Tiehui~Ted~Liu, 
  Hang~Yin
\thanks{Manuscript received November 22, 2013. This work was supported by the US Department of Energy Office of Science.}
\thanks{Y.~Okumura is with University of Chicago, Chicago, Illinois 60637, USA and Fermi National Accelerator Laboratory, Batavia, Illinois 60510, USA. (e-mail: Yasuyuki.Okumura@cern.ch).}%
\thanks{J.~Olsen, T.~Liu and H.~Yin are with Fermi National Accelerator Laboratory, Batavia, Illinois 60510, USA.}%
}

\maketitle
\thispagestyle{empty}


\begin{abstract}
High luminosity conditions at the LHC pose many unique challenges for potential silicon based track trigger systems.
One of the major challenges is data formatting, where hits from thousands of silicon modules must first be shared and organized into overlapping eta-phi trigger towers.
Communication between nodes requires high bandwidth, low latency, and flexible real time data sharing, for which a full mesh backplane is a natural solution.
A custom Advanced Telecommunications Computing Architecture data processing board is designed with the goal of creating a scalable architecture abundant in flexible, non-blocking, high bandwidth board to board communication channels while keeping the design as simple as possible.
We have performed the first prototype board testing and our first attempt at designing the prototype system has proven to be successful.
Leveraging the experience we gained through designing, building and testing the prototype board system we are in the final stages of laying out the next generation board, 
which will be used in the ATLAS Level-2 Fast TracKer as Data Formatter, as well as in the CMS Level-1 tracking trigger R\&D for early technical demonstrations.
\end{abstract}

\IEEEpeerreviewmaketitle

\section{Introduction}
\IEEEPARstart{H}{igh} luminosity conditions at the LHC pose many unique challenges
for potential silicon based track trigger systems. This is true
for both Level-1 and Level-2 trigger applications. Among those
data formatting is one of major challenges, where hits and clusters 
from many thousands of silicon modules must first be shared and 
organized into overlapping eta-phi trigger towers due to finite 
size of the beam’s luminous region along the beam axis and 
the finite curvature of charged particles in the magnetic field. 
Communication between nodes requires high bandwidth, low latency, and flexible real time
data sharing. The first silicon based track trigger at the LHC will
be the ATLAS Fast Tracker (FTK) at Level-2~\cite{FTK_TDR}.
Although ATLAS FTK is designed for Level-1 Accept rates up to 100~kHz,
the data volume per event is quite large since all silicon
modules (more than 86 million channels) are involved at high 
luminosity, therefore this is where challenging data formatting 
issues will be encountered for the first time. We have been 
developing data formatting solutions for high luminosity LHC 
conditions and the ATLAS FTK Data Formatter system~\cite{DF_SPEC} 
is the first targeted application.

\section {ATLAS FTK Data Formatter}

\begin{figure}[ht!]
  \centering
  \includegraphics[width=0.45\textwidth,clip]{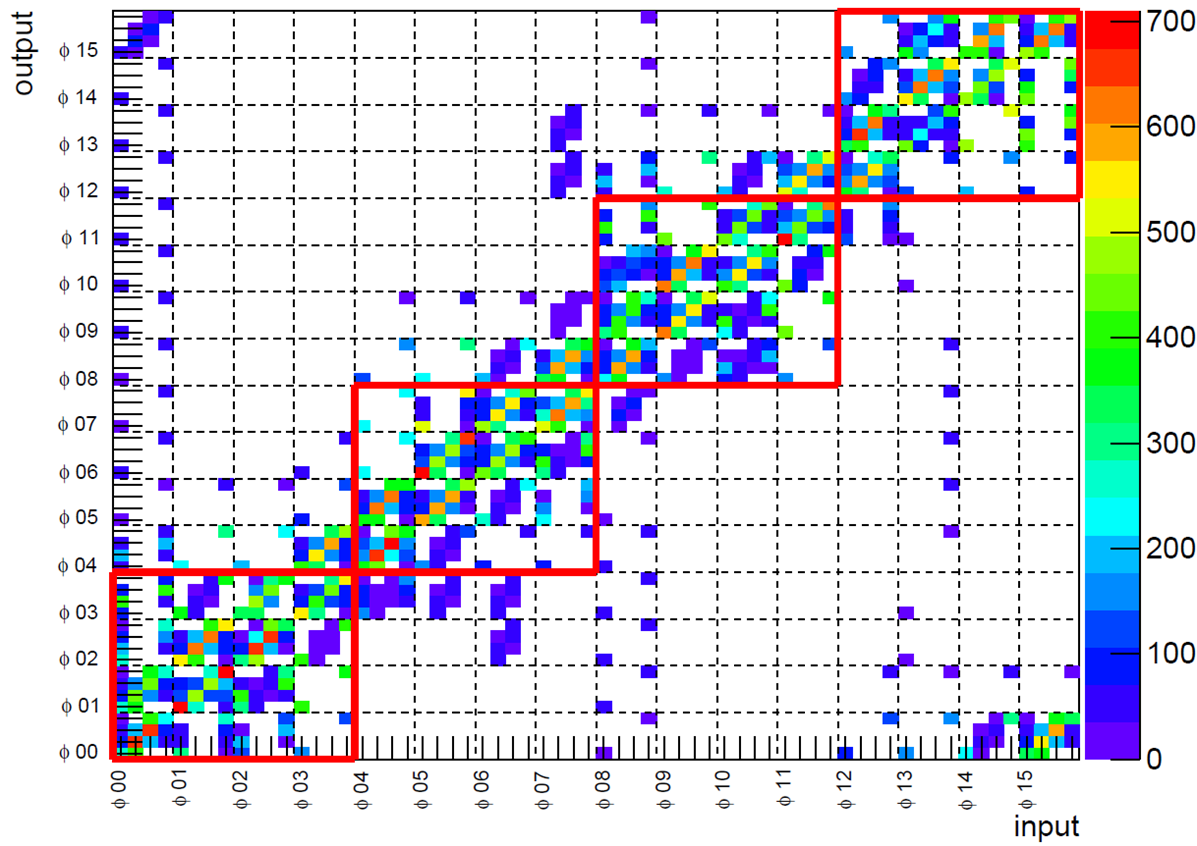}
  \caption{The  $64 \times 64$ matrix shows the required data sharing in ATLAS FTK
    Data Formatter among 64 eta-phi trigger towers. The four red boxes indicate the
    assignment of trigger towers to four crates to minimize inter-crate data sharing.
    The color scale indicates the number of clusters shared between trigger towers 
    per event evaluated with LHC-ATLAS data taken in 2012~\cite{DF_SPEC}.}
  \label{fig:input_output}
\end{figure}

\begin{figure}[ht!]
  \centering
  \includegraphics[width=0.35\textwidth,clip]{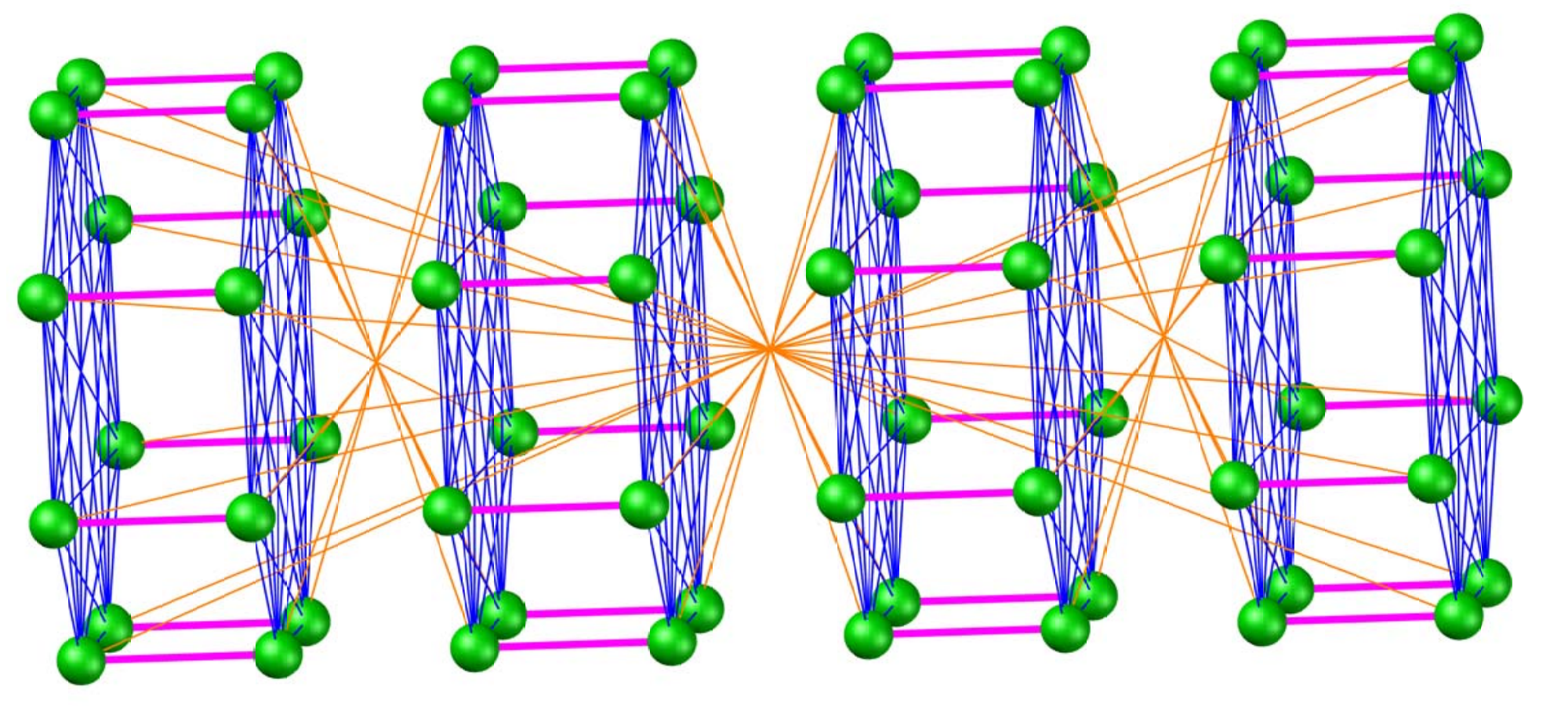}
  \caption{A graphical depiction of the 64 trigger towers (in green) and high speed interconnect lines in four crate system.  
Blue lines represent backplane data paths.  A high speed local bus is shown in purple.  Orange lines represent inter-crate fiber links.}
  \label{fig:tower_connection}
\end{figure}

The ATLAS FTK is organized as a set of parallel processor units
within an array of 64 eta-phi trigger towers.
The $64 \times 64$ matrix in Figure~\ref{fig:input_output} shows
the required data sharing in the data formatting stage first among the ATLAS FTK
eta-phi trigger towers.
Because the existing silicon tracker and front end electronics were not designed for triggering, the data sharing among trigger
towers is quite complex, as shown in the matrix.
The Data Formatter hardware design is dominated by the input and output requirements, and we analyzed
the data sharing in early design stage  using real beam data with the actual readout cable mapping. 
The four red boxes in the matrix
represent crate boundaries.  Boards within each crate communicate over the backplane.
Fiber links are used when boards must communicate across crate boundaries.
Our analysis shows that the data sharing between trigger towers is highly dependent
upon upstream cabling and detector geometry.  The ideal Data Formatter hardware platform 
should be flexible enough to accommodate future expansion and allow for changes in input 
cabling and module assignments.  One example of such a system is shown in Figure~\ref{fig:tower_connection}, where 
each trigger tower is represented by a green ball and lines represent data paths.

\section {Design Concept}

\begin{figure}[ht!]
  \centering
  \includegraphics[width=0.30\textwidth,clip]{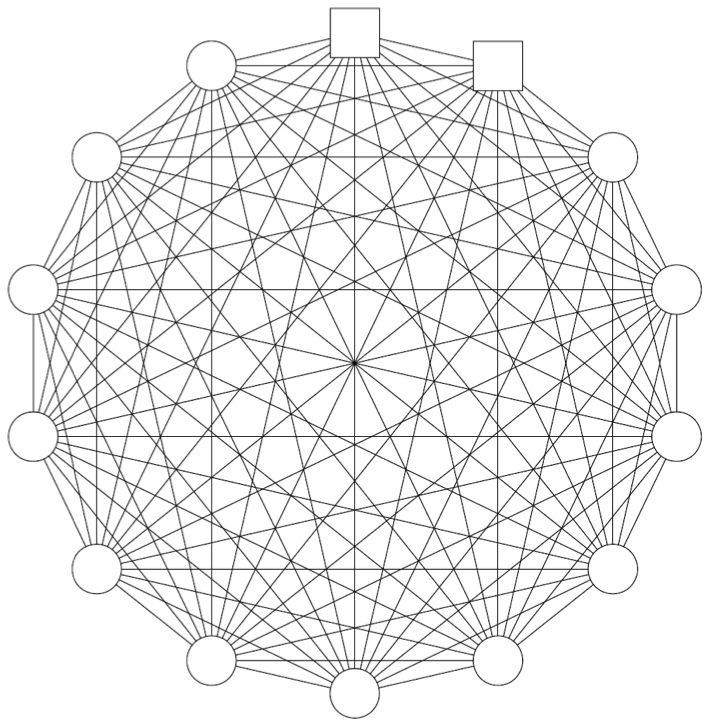}
  \caption{The fabric interface connections in 14 slot full mesh ATCA backplane.  Each line represents a multi-lane bidirectional channel rated for up to 40~Gb/s.}
  \label{fig:fullmesh}
\end{figure}

Our hardware design process followed a bottom up approach whereby
we studied various track trigger architectures. Implementations involving 
full custom backplanes and discrete cables were considered.  Eventually 
the full mesh Advanced Telecommunication
Computing Architecture (ATCA)~\cite{picmg30} backplane was found to
be a natural fit for the Data Formatter requirements. The Fabric
Interface of the full mesh backplane enables high speed point-to-point
communication between every slot (Figure~\ref{fig:fullmesh}),
with no switching or blocking. Each line in this diagram represents a channel which consists
of up to four bidirectional lanes, which runs at the maximum speed of 40~Gb/s.
Field Programmable Gate Array (FPGA) devices, which are abundant in local cells, memory, and high speed serial transceivers,
were selected for the core processing element on each Data Formatter board.


\begin{figure}[ht!]
  \centering
  \subfigure[]{
    \includegraphics[width=0.20\textwidth,clip]{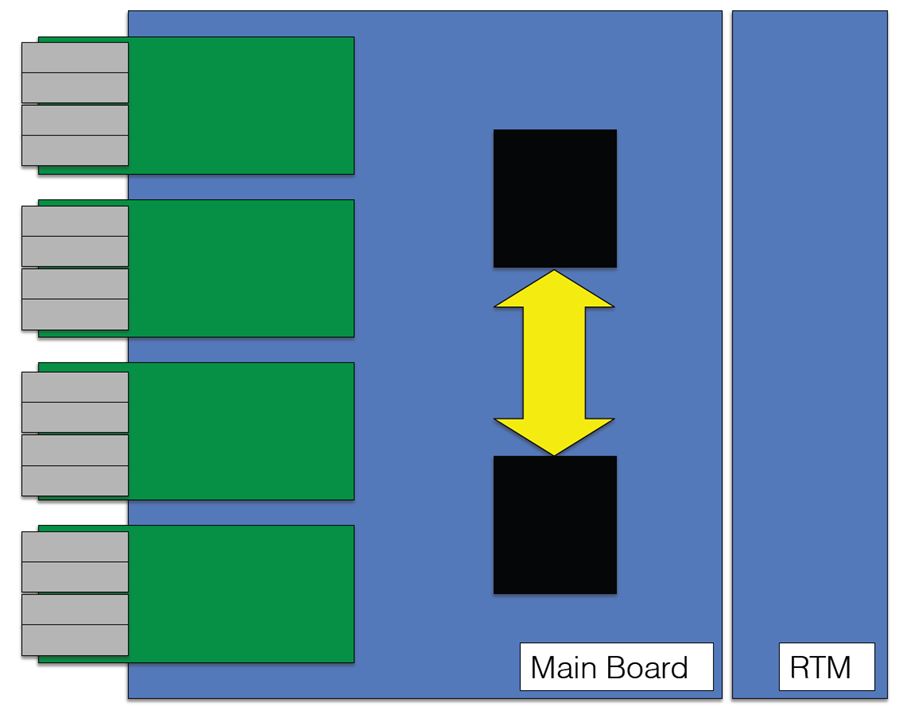}
    \label{fig:local_bus_interface}
  }
  \subfigure[]{
    \includegraphics[width=0.15\textwidth,clip]{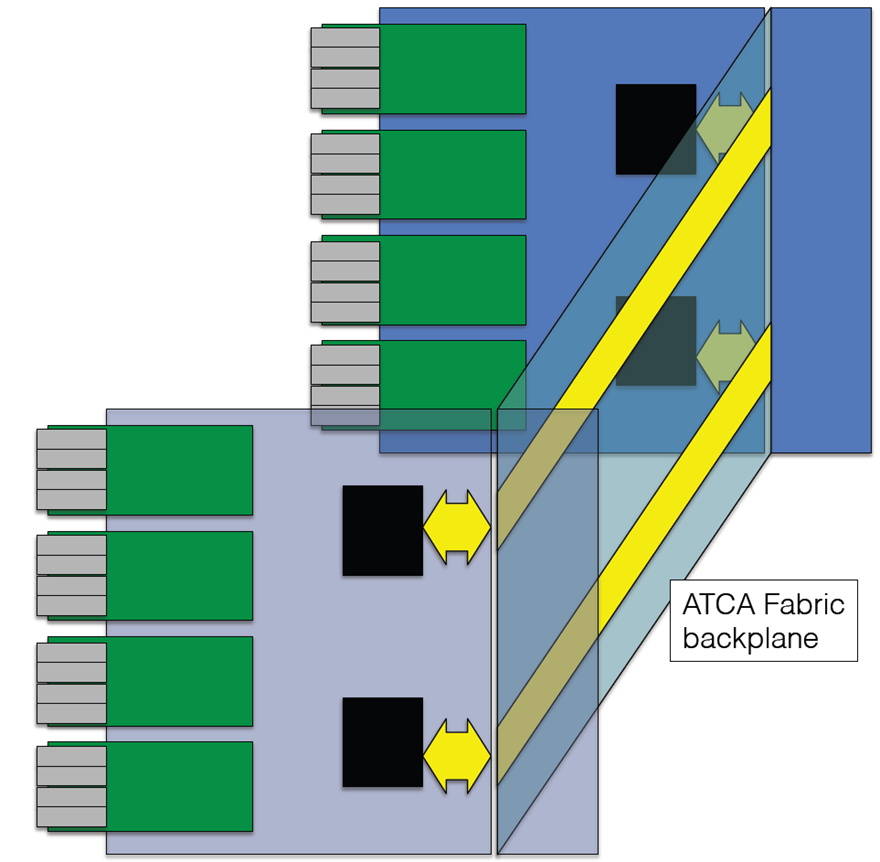}
    \label{fig:atca_fabric_interface}
  }
  \subfigure[]{
    \includegraphics[width=0.35\textwidth,clip]{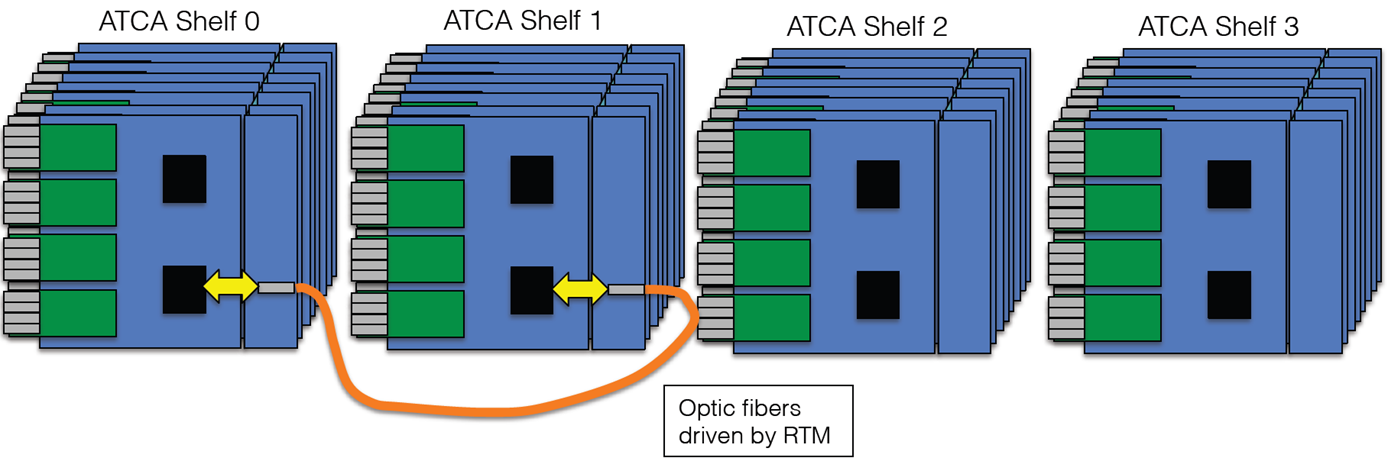}
    \label{fig:optical_driver_interface}
  }
  \caption{The FPGAs are interfaced to (a) local bus connecting
    two FPGAs on the board, (b) ATCA full mesh backplane for use of
    point-to-point links and (c) pluggable transceivers on RTM.
    }
  \label{fig:internal_connectity}
\end{figure}

High speed serializer components in the FPGA are directly connected to the full mesh
backplane fabric interface channels and also to pluggable fiber transceivers
located on a rear transition module (RTM).  Our first prototype ATCA board incorporated 
a pair of FPGAs, and thus required a high speed local bus to implement the three types of 
interconnects described in Figure~\ref{fig:internal_connectity}.

The full mesh backplane, when coupled with large FPGAs rich in high speed
serial transceivers, offers a level flexibility previously unattainable in a
conventional shared bus architecture.  In effect, the system blurs the
distinction between individual FPGAs and dramatically increases the processing
power and I/O capability.

\section {The Pulsar IIa Prototype}
\label{sec:pulsar2a}

\begin{figure}[ht!]
  \centering
  \includegraphics[width=0.45\textwidth,clip]{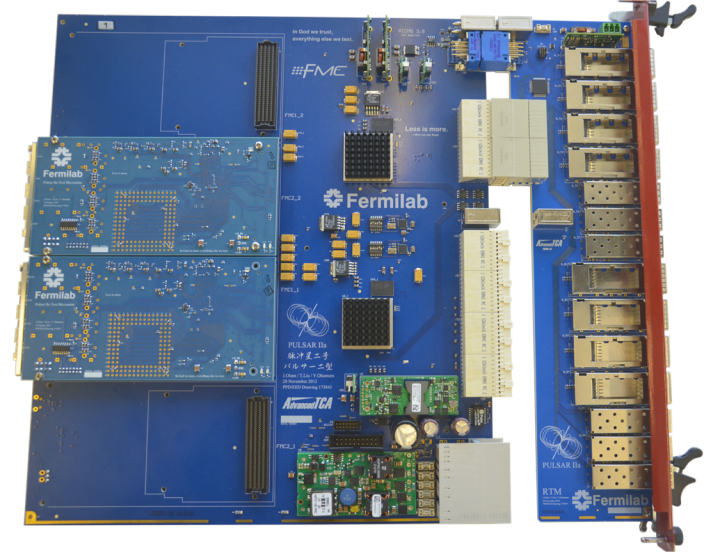}
  \caption{The Pulsar IIa prototype ATCA front board and rear transition module.}
  \label{fig:pulsar2a_system_photo}
\end{figure}

We designed the first prototype, the Pulsar IIa, to gain experience using the
latest FPGAs with their high speed transceivers in an ATCA form factor.  
The Pulsar IIa consists of a front board and rear transition module (RTM)
as shown in Figure~\ref{fig:pulsar2a_system_photo}.  In addition we developed 
an FMC~\cite{fmc} mezzanine card which includes a smaller FPGA, four fiber transceivers, 
and a socket used for ASIC testing.  A mini-backplane was also developed to 
facilitate standalone board testing on the bench top.  This section describes
the design details of each component.  Further design details also can be accessed on our open
web site~\cite{atca_at_fermilab}.

\subsection {Pulsar IIa Front Board}

\begin{figure}[ht!]
  \centering
  \includegraphics[width=0.45\textwidth,clip]{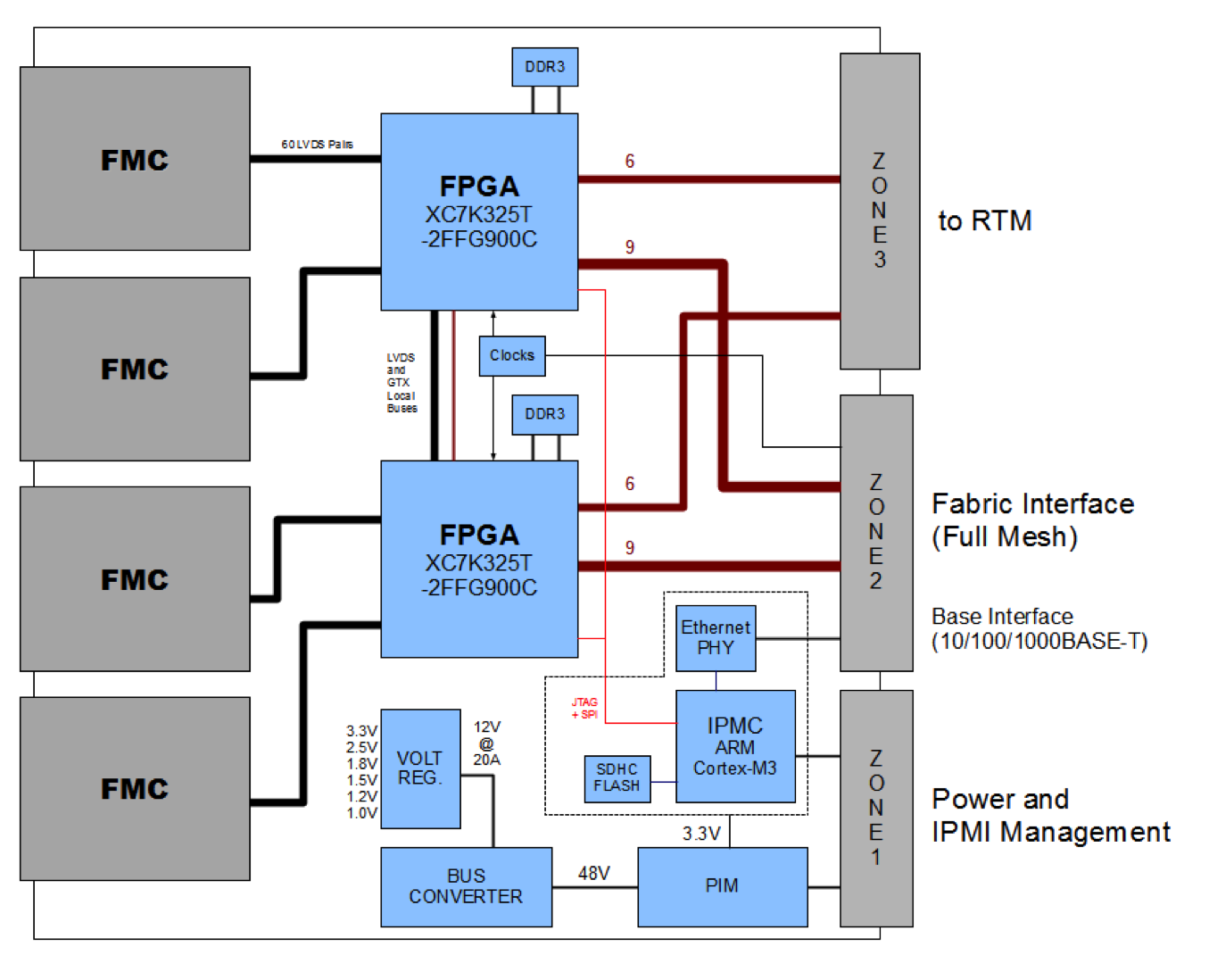}
  \caption{The Pulsar IIa prototype board block diagram.  Four FMC mezzanine cards are supported.}
  \label{fig:proto_block}
\end{figure}

The Pulsar IIa front board is designed around a pair of Xilinx Kintex XC7K325T
FPGAs.  The block diagram of the board is shown in Figure~\ref{fig:proto_block}.
Each FPGA has 16 high speed serial transceivers (GTX) which support data rates up to
10~Gb/s.  Of these 16 GTX transceivers 9 connect to the fabric interface, 6
connect to the RTM, and the last GTX transceiver is used for the local bus.

Each board supports up to four high pin count FMC mezzanine cards, which are
connected to the main FPGAs using general purpose LVDS signal pairs.  The FMC form factor
has in recent years become a popular choice with Xilinx development boards and
many third party developers.  When the Pulsar IIa board is used as the FTK Data Formatter
four cluster finder input mezzanine cards will be used.

A Cortex-M3 microcontroller is used as an
Intelligent Platform Management Controller (IPMC),
which is required on all ATCA boards. This microcontroller
is responsible for the following:
\begin {itemize}
\item Implementing the IPMI protocol and communicating with the ATCA shelf manager board(s), coordinating hot swap operations, etc.
\item Running Telnet and FTP servers which are accessed via the 100BASE-T Base Interface Ethernet port.
\item Managing firmware images on a micro SDHC flash card.
\item Programming the FPGAs via JTAG and monitoring over an SPI bus.
\item Reading various board temperature and voltage sensors.
\item Communicating with the RTM over an I$^{2}$C bus.
\end {itemize}

\subsection {Rear Transition Module}

The Pulsar IIa RTM conforms to the PICMG3.8 standard~\cite{picmg38} and is
considered an intelligent-FRU device.  A small ARM microcontroller on the RTM
continuously monitors the status of the 8 QSFP+ and 6 SFP+ pluggable
transceivers.  The microcontroller also communicates with the front board IPMC
microcontroller and coordinates hot swap sequencing.  Each of the Pulsar IIa
FPGAs connects to one QSFP+ transceiver and two SFP+ transceivers on the RTM.

\subsection {Mezzanine Card}

\begin{figure}[ht!]
  \centering
  \includegraphics[width=0.45\textwidth,clip]{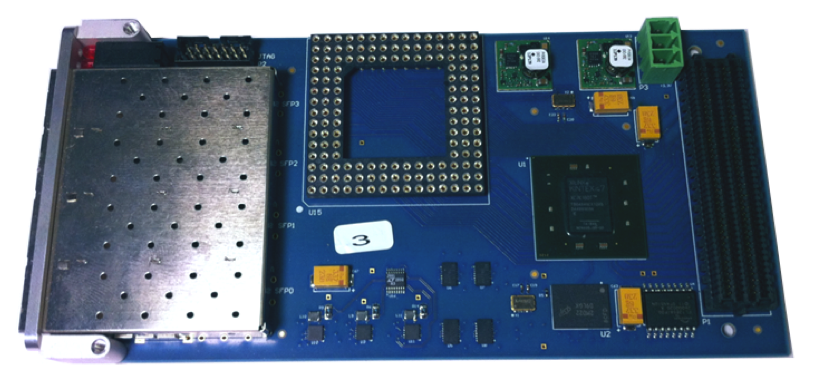}
  \caption{The test mezzanine card with FMC connector, SFP+ transceivers, and an ASIC test socket.}
  \label{fig:test_mezzanine_photo}
\end{figure}

The Pulsar IIa supports up to four FMC mezzanine cards with the high pin count 
(HPC) LVDS interface.  Mezzanine cards may contain FPGAs, pattern recognition
ASICs, fiber optic transceivers, or any other custom hardware.  We developed
our FMC test mezzanine card, shown in Figure~\ref{fig:test_mezzanine_photo}, in
order to become familiar with the FMC form factor and to study high speed LVDS
communication between FPGAs.  The test mezzanine card features a Xilinx Kintex
XC7K160T FPGA, 4 SFP+ transceivers, 128MB DDR3, and a 144 pin socket used for
testing custom ASIC chips, primarily aimed at testing pattern recognition
associative memory devices~\cite{VIPRAM}.  The FMC connector supports 3.3V and
12V power.  An I$^2$C bus and JTAG interface are also provided for slow controls
and in-system programming.

\subsection {Mini Backplane}

\begin{figure}[ht!]
  \centering
  \includegraphics[width=0.40\textwidth,clip]{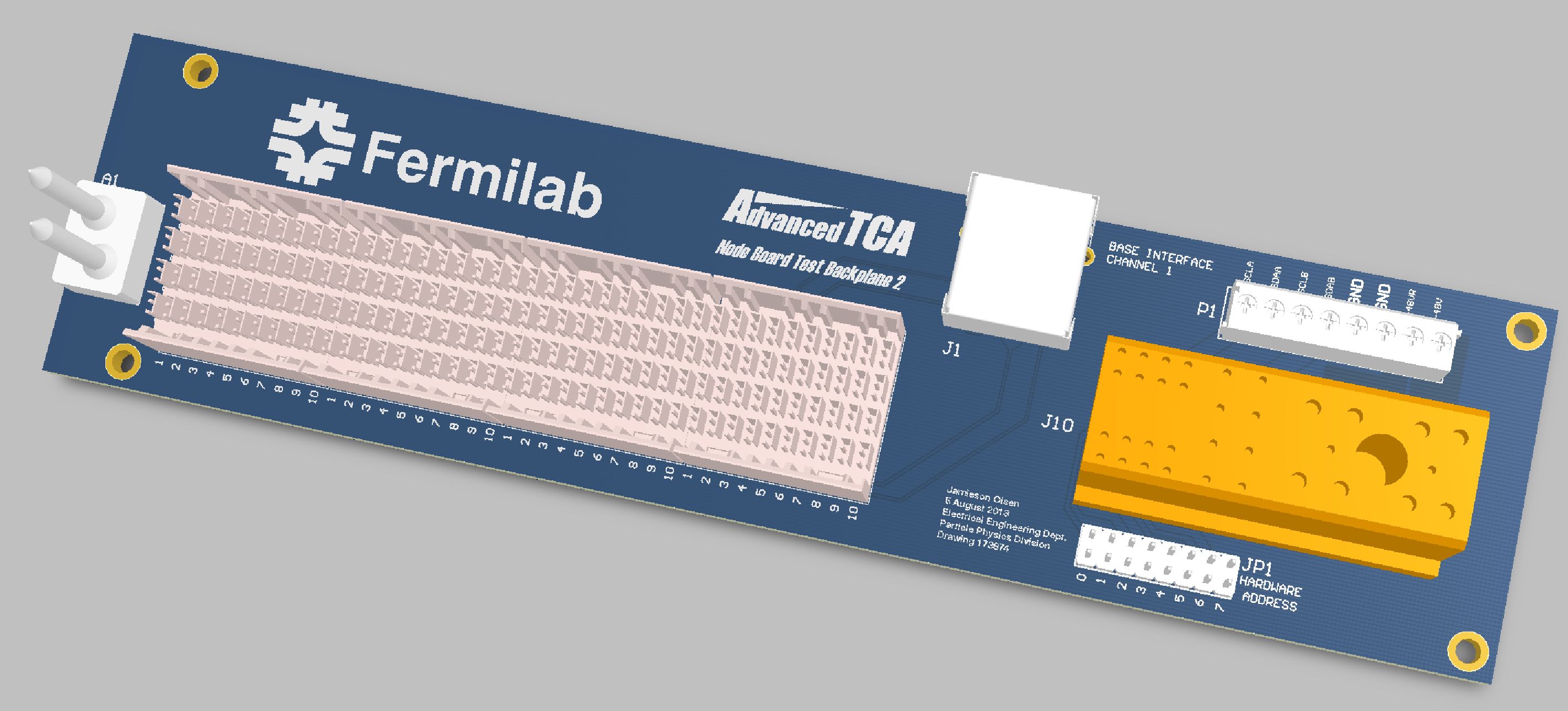}
  \caption{The mini backplane and a 48VDC power supply are all that is required to power up the Pulsar IIa board on the bench.}
  \label{fig:mini_backplane}
\end{figure}

The Mini Backplane (Figure~\ref{fig:mini_backplane}) was developed to support
stand alone testing of a single Pulsar IIa board and RTM on the bench top.  Power is
supplied from a 48VDC bench supply.  The Base Interface Ethernet port is brought out to
an RJ45 connector which may then be connected to a PC and used for
communicating with the IPMC microcontroller.

The Mini Backplane also loops back each Fabric Interface port.  This loopback feature has proved
to be very useful in testing and characterizing the performance of the GTX
serial transceivers outside of the crate.

\section {Test Results}

\subsection {Board-level testing}

\begin{figure}[ht!]
  \centering
  \includegraphics[width=0.40\textwidth,clip]{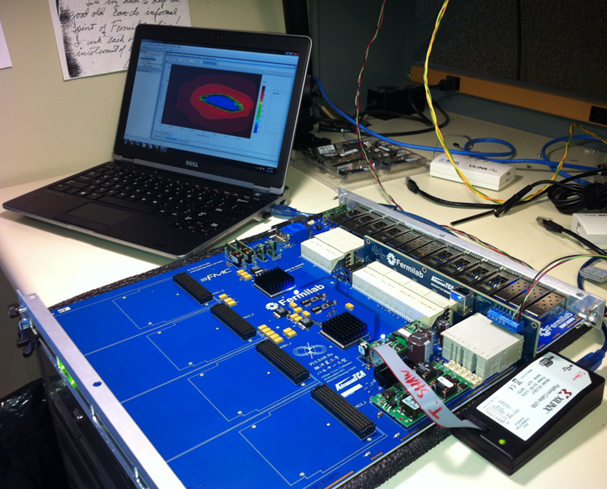}
  \caption{The mini backplane powers the Pulsar IIa board on the bench top.
    A laptop and 48VDC power supply is all that is required to power up and test the Pulsar IIa board and RTM.}
  \label{fig:board_level_testing_photo}
\end{figure}

Figure~\ref{fig:board_level_testing_photo} shows our bench top test setup,
which is used to check basic functionality such power supply operation, IPMC
microcontroller communication, FPGA programming and GTX loopback tests.

The FPGA GTX transceivers are driving data in the test-stand system
in the Fabric Interface channels, the RTM channels and the Local Bus.
It turns out that the FPGAs achieved a bit error rates (BER)
less than $10^{-16}$ level in long-running loop back tests.
The upper limit of BER is defined to be $1/{total~number~of~transported~bits}$
if there is no single error detected during the test.
Table~\ref{tab:long_hour_test} summarized the measured upper limit.

\begin{table}[ht!]
  \centering
  \begin{tabular}{|l||c|c|} \hline
    & Speed    & BER upper limit \\ \hline
    Fabric Interface channels & 6.25~Gb/s & $4.2\times10^{-17}$ \\ \hline
    RTM channels              & 6.25~Gb/s & $8.3\times10^{-17}$ \\ \hline
    Local Bus                 & 10~Gb/s   & $1.4\times10^{-15}$ \\ \hline
  \end{tabular}
  \caption{BER upper limit measurement results. The upper limit depends on the
    length of the long hour test for each channel. }
  \label{tab:long_hour_test}
\end{table}

\begin{figure}[ht!]
  \centering
  \includegraphics[width=0.40\textwidth,clip]{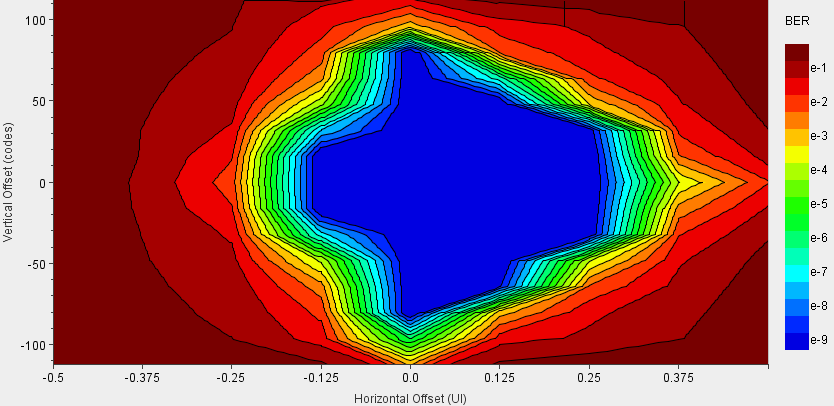}
  \caption{An example of measured statistical eye. This is
    a result for local bus GTX transceiver channel at 10~Gb/s.}
  \label{fig:eye_diagram}
\end{figure}

In addition to the BER test, we performed a receiver margin analysis, or eye scan, using the Xilinx IBERT tool.
The Kintex-7 GTX transceivers have built-in diagnostic features which
provide a mechanism to measure and visualize the receiver eye
margin after RX equalizer~\cite{gtx}. Sweeping the receiver sampling point and vertical 
offset voltage enables the generation of a BER map or statistical eye diagram, where the color
represents $\log_{10}({\rm BER})$.
Figure~\ref{fig:eye_diagram} shows an example of the measured statistical eye diagram
and the open blue region indicates that we can have error
free operating points. The size of eye corresponds to the quality
of the high speed serial communication after the RX equalization.
All GTX transceiver channels have been tested and characterized using the IBERT tool.
Furthermore, the eye scan has been done with a Xilinx Kintex-7 evaluation
kit (KC705)~\cite{kc705}, which provides a "golden" reference for comparison studies.
Comparing the Pulsar IIa eye diagrams against the Xilinx reference design helps
us learn more about high speed layout techniques, which will be used in the
next iteration of the board (Section~\ref{sec:pulsa2b}).

Communication over the LVDS lines between the FMC mezzanine and the main FPGAs
has been tested successfully at 400~MHz single data rate (SDR) and 200~MHz double
data rate (DDR).  Thirty-four LVDS pairs running at this speed yield a
bandwidth of 13.6~Gb/s, which exceeds the ATLAS FTK requirements for the Data
Formatter board.

\subsection {Crate-Level Testing}
\begin{figure}[ht!]
  \centering
  \includegraphics[width=0.35\textwidth,clip]{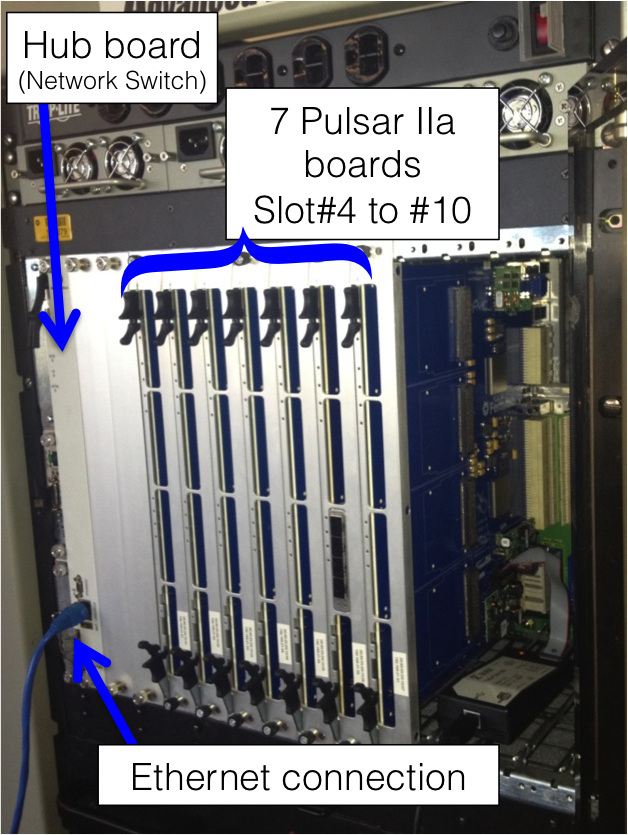}
  \caption{Our 14 slot ATCA shelf with a commercial switch blade in slot 1 and seven Pulsar IIa boards in slots 4 through 10.}
  \label{fig:crate_level_testing_photo}
\end{figure}

A crate-level testing followed the single board-level testing.
We set up an ATCA shelf with a network switch blade in the hub slot
and seven Pulsar IIa front boards as well as RTMs in the node slots,
as shown in Figure~\ref{fig:crate_level_testing_photo}.

In our system test the IPMC microcontroller communicates with the switch blade
over the Base Interface network.  We first log into the switch and then telnet
into each Pulsar IIa board, where commands are issued to program the FPGAs and
monitor various board sensors.  The default firmware image is setup to drive
data over all GTX transceivers to the fabric, RTM and local bus channels.  

\begin{figure}[ht!]
  \centering
  \includegraphics[width=0.40\textwidth,clip]{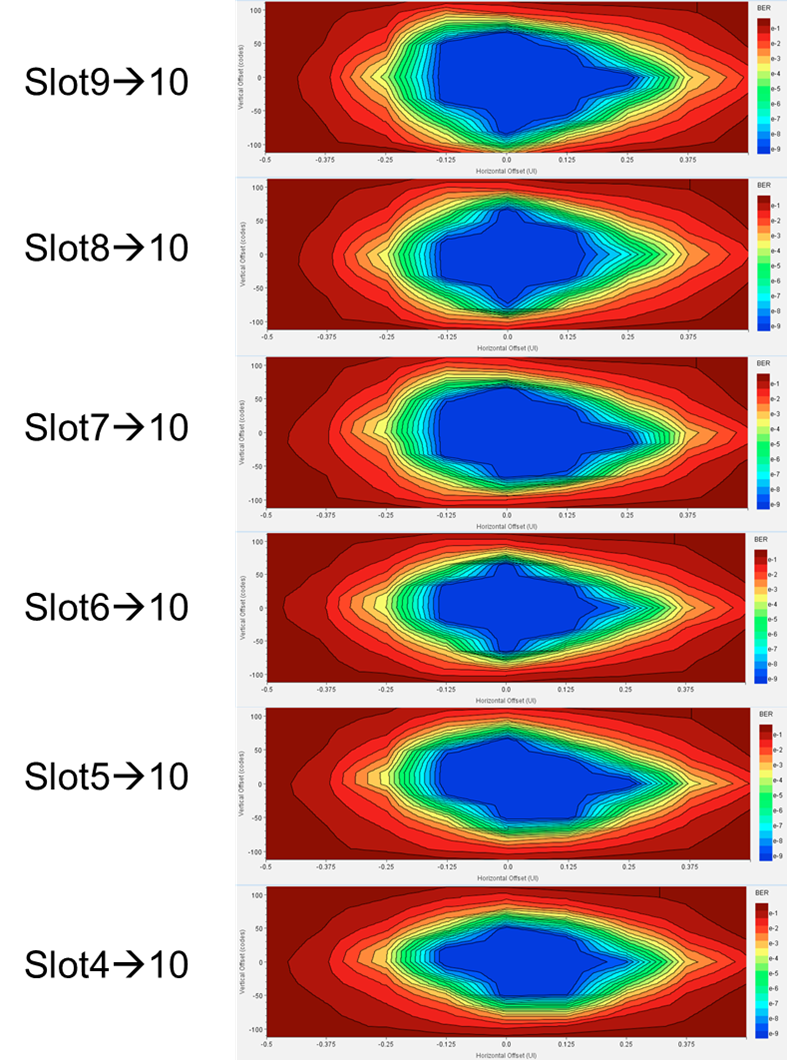}
  \caption{A set of statistical eyes measured in ATCA crate level operation.
The receiving board is in slot 10 while the location of the transmitting
board varies.  At this bit rate performance degradation along the length of the backplane is negligible.}
  \label{fig:eye_scan_ATCA}
\end{figure}

The Xilinx IBERT tool has also been used in the the crate to test the backplane
performance.  While our 10G ATCA backplane is rated for only 3~Gb/s per lane it
has proven to work admirably at up to 6.25~Gb/s across all slots, as shown in
Figure~\ref{fig:eye_scan_ATCA}.

\section {Pulsar IIb}
\label{sec:pulsa2b}
\begin{figure}[ht!]
  \centering
  \includegraphics[width=0.45\textwidth,clip]{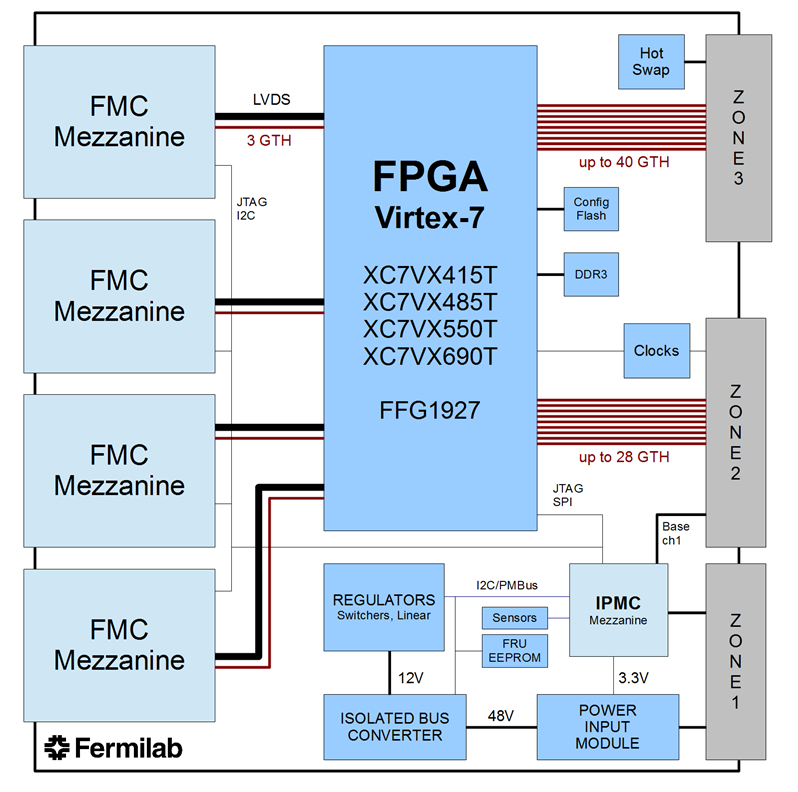}
  \caption{The Pulsar IIb block diagram. The IPMC microcontroller and associated circuitry are now located on a small mezzanine card.}
  \label{fig:pulsariib_block}
\end{figure}

\begin{figure}[ht!]
  \centering
  \includegraphics[width=0.40\textwidth,clip]{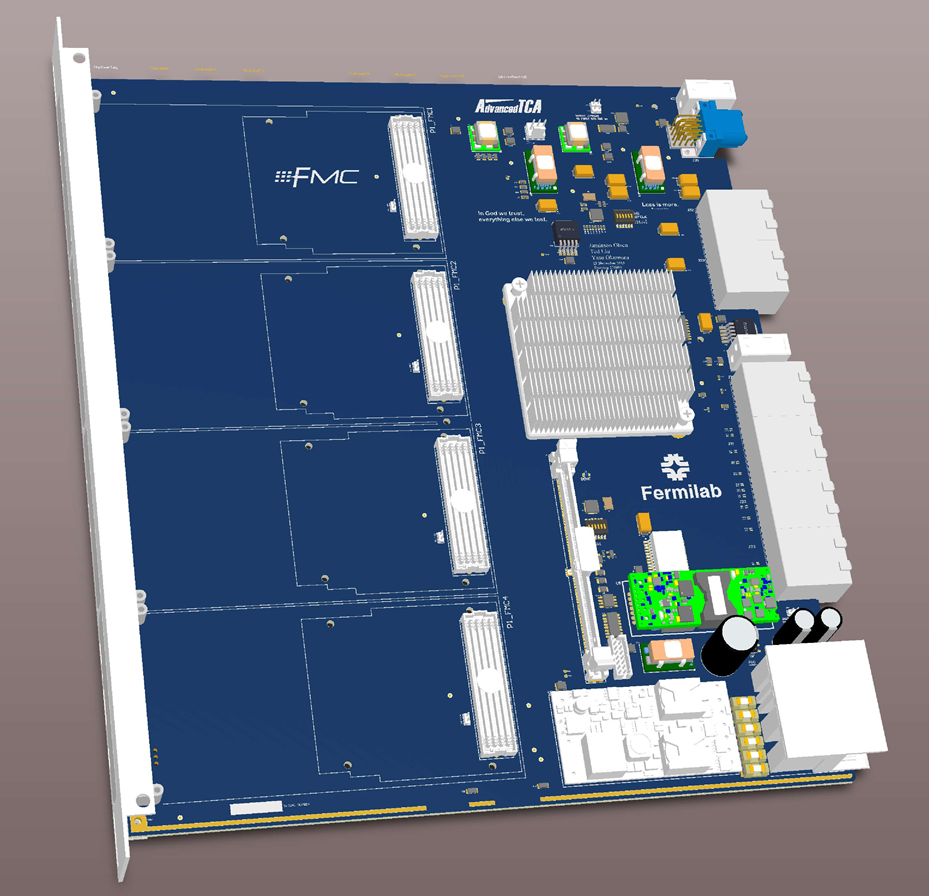}
  \caption{A 3D view of the Pulsar IIb layout.}
  \label{fig:pulsariib_photo}
\end{figure}

Leveraging the experience we gained through designing,
building and testing the Pulsar IIa system we are in the final stages
of laying out the next generation board, the Pulsar IIb
(Figure~\ref{fig:pulsariib_block} and Figure~\ref{fig:pulsariib_photo}).
The new board design replaces the two Kintex K325T devices
with a single large Virtex-7 FPGA.  The GTX transceiver
count has increased up to 80 channels, providing a significant
bandwidth increase to the RTM, Fabric and Mezzanine cards.
The power regulator sections of the board have been redesigned to
handle the increased power required by the Virtex-7 FPGA.

The Pulsar IIb boards will be used for the ATLAS FTK Data
Formatter system.  We anticipate that the boards will also
be used for CMS L1 tracking trigger early technical 
demonstrations~\cite{TWEPP2013_PAPER}.

\section {Conclusion}

The Pulsar IIa is our first ATCA prototype board and works as designed, as
demonstrated by our successful stand-alone and crate-level tests.  Through this
prototype development process we have gained experience using the latest Xilinx
FPGAs and high speed serializers to communicate over the ATCA full mesh
backplane.  Furthermore, the Pulsar IIa boards have successfully interfaced with other ATCA
system components such as commercial switch blades and shelf manager cards.


The Pulsar IIb boards will be used in the ATLAS FTK Data Formatter system
starting in 2015. The Pulsar IIb design forms the basic building block of a high performance scalable architecture, which may find
applications beyond tracking triggers, and may serve as a starting point for
future Level-1 silicon-based tracking trigger R\&D for CMS,
where the full mesh backplane is used most effectively for sophisticated time multiplexing data transfer schemes.


Our baseline design also works well as a general purpose FPGA-based processor board.
The design may prove useful in scalable systems where highly flexible,
non-blocking, high bandwidth board to board communication is required.

\end{document}